# Effect of Functional Group on Electrical Switching Behaviour of an Imidazole Derivative in Langmuir-Blodgett Film

Bapi Dey[a], Sudip Suklabaidya[a], Swapan Majumdar[b], Pabitra Kumar Paul[c], Debajyoti. Bhattacharjee[a], Syed Arshad Hussain*[a]

[a]Thin film and Nanoscience Laboratory, Department of Physics, Tripura University, Suryamaninagar 799022, Tripura, India

[b]Department of Chemistry, Tripura University, Suryamaninagar 799022, Tripura, India

[c]Department of Physics, Jadavpur University, Kolkata - 700032, West Bengal, India.

* Corresponding author

Email: sa_h153@hotmail.com, sahussain@tripurauniv.in

Ph: +919402122510 (M), +91381 2375317 (O)

Fax: +913812374802 (O)

Here we report the design and synthesis of an imidazole derivative namely 1-benzyl-2,4,5-triaryl imidazole (compound **2**) and its switching behaviour assembled onto Langmuir-Blodgett (LB) films. Monolayer characteristic of compound **2** at the air- water interface has been studied by surface pressure vs area per molecule ($\pi$ – A) isotherm, hysteresis analysis and in-situ Brewster Angle Microscopy (BAM). These studies indicated the formation of stable floating Langmuir film at the water subphase. Atomic Force Microscopy (AFM) investigation confirmed the successful deposition of the Langmuir film onto solid substrate. Device consisted of 60 layers LB films of **2** showed resistive bipolar switching behaviour irrespective of the first applied bias voltage polarity. Observed bipolar switching has been explained in terms of reduction – oxidation process. Due to the presence of strong reducible group (C = N) in the imidazole core, reduction – oxidation process takes place easily during bias. Presence of sharp reduction and oxidation peaks in the Cyclic Voltammetry (CV) measurement of **2** also supported this hypothesis. Presence of benzyl group with the imidazole core played the crucial rule in the reduction – oxidation process and hence the switching behaviour. When benzyl group was replaced by a '–H' then bipolar switching was not observed. In that case oxidizable group N – H opposed the reduction process during bias. This type of bipolar switching is very promising for future technological applications in organic electronics.

# 1. Introduction

Electronic and optoelectronic devices are the basic building blocks for the modern civilization.[1–7] Recent extensive studies have shown that distinctive advantages of organic materials, such as low cost of processing, lightweight, mechanical flexibility, ambient processing, printability, variety of materials etc. has attracted specific attention towards organic electronics.[8–16] Organic electronics deals with various interesting organic materials and the innovative organic nanomaterials exhibit a variety of important properties such as optical, electrical, photoelectrical conducting, semiconducting, memory, storage and magnetic properties.[17–21] In addition, several organic materials have been used as the principle component to design various electronic and optoelectronic devices, such as diodes, sensors, organic light-emitting diodes (OLED), organic field effect transistors (OFET), solar cells, lasers, detectors, memory-switching devices, logic gates etc.[22–28] Nowadays, molecular electronics has emerged as an important technology. It deals with the design, processing and device application of organic molecules at the molecular level/ nanoscale level.[29–31] As the industry moves from bulk to molecular electronics, there is a growing trend to revisit voltage-induced switching phenomena in conjugated organic materials, which was initially observed more than thirty years ago.[32,33] Switching devices fabricated by using organic molecules are the promising candidates for the next generation of non-volatile memories due to their simple structure, low cost, excellent performance and great scale-down potential.[34–41] By applying a suitable bias voltage a switching device can be switched between two states, a low conducting OFF state and a high conducting ON state.[42–46] Depending on the ability to retain information, switching behaviour can be classified into two types – volatile and non-volatile (memory) switching. In case of memory switching it is possible to retained both ON and OFF state even after the removal of externally applied bias voltage, whereas in threshold switching only OFF state is stable at lower bias voltage.[47] Based on the molecular structure, conformation, and types of contacts, this types of two conducting state switching behaviour of various organic materials in ultrathin films has already been observed during current–voltage (I–V) characteristics measurement for metal/molecular thin film/metal structures.[48–55] Therefore suitable organic molecules for designing switching devices, which have found their potential application in logic and memory circuits are the subject of current interest. Since the past several years, enormous initiative has been taken for the design and characterization of such innovative organic materials having a readily polarizable structure to fabricate novel devices.[56–59] Imidazole derivatives having π – conjugated backbone and two nitrogen atoms of distinct electronic property are very promising in this regard. It is possible to form donor– acceptor system when different aromatic groups are added to them. Accordingly imidazole derivatives can show interesting electrical and optical properties.[60–63]. A series of imidazole derivatives incorporating thiophene and polyaromatic hydrocarbon groups have been designed and synthesized for organic light emitting device application.[61,64,65]. It has been observed that imidazole derivative functionalized with fullerene and fluorene can be used for solar cell application.[66-67] Electrochemical sensor based on imidazole functionalized grapheme oxide has also been demonstrated.[68] Imidazole incorporated onto phtalocyanines has been synthesized for electrical and gas sensing applications.[69] Chemosensor and bright emission tuneable dyes based on different imidazole derivatives have also been reported.[62,70] Nano to micro dimensional thin films of organized molecular arrays are the prerequisite for the design of molecular and electronic devices.[71,72] LB method is very promising and unique in this regard to prepare such thin films, where it is possible to control/manipulate the molecular arrangement in such films.[73–75]

Recently, we have designed and synthesized an interesting imidazole derivative (abbreviated as **1**).[71] LB technique was used to investigate its assembly behaviour onto solid support.[71] Molecule **1** formed supramolecular nanostructures when assembled onto LB films and showed electrical threshold switching in LB films. However, no bipolar (memory) switching was observed. Intermolecular charge transfer within molecule **1** was mainly responsible for such observed switching behaviour.[71] For compound **1** reduction process was unlikely due to the presence of oxidizable group (N – H) in the imidazole core (Figure 1) although reduction – oxidation process is key for bipolar memory switching.[76,77] Therefore, we have planned to replace the – H of NH functionality located in imidazole ring of **1** by a benzyl group (compound **2**, present study). Accordingly we have synthesized compound **2** (Figure 1) and investigated its memory switching behaviour. We expected that in presence of benzyl group reduction – oxidation process would occur within the molecule **2**, which would lead to the bipolar memory switching under suitable bias condition. Interestingly our investigation revealed that device based on compound **2** showed resistive bipolar switching behaviour in LB films. This kind of bipolar switching will play crucial role in future organic electronic devices especially in memory application.[78]

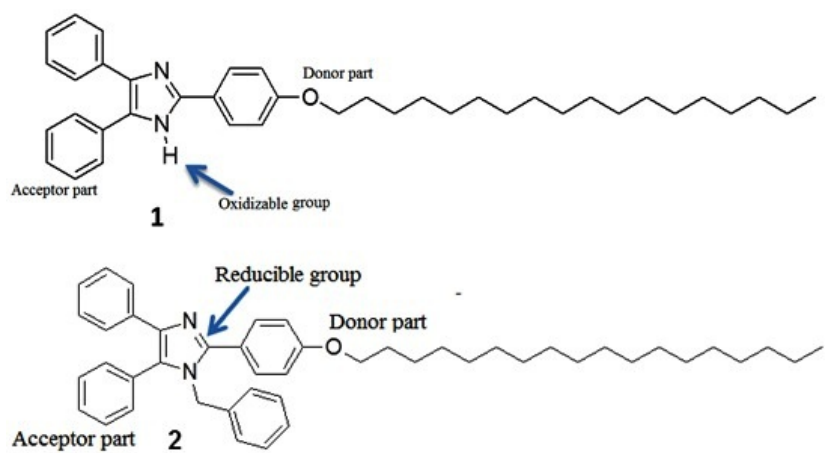

**Figure 1.** Chemical structure of 2,4,5-triaryl imidazole derivative (**1**) and 1-benzyl-2,4,5-triaryl imidazole derivative (**2**).

## 2. Experimental Results

### 2.1. Monolayer characteristics at the air-water interface

To have an idea about the thermodynamic behaviour of **2**, π−A isotherm of **2** spread onto the water subphase has been measured is shown in Figure 2.

Isotherm curve (Figure 2) started to rise with initial lift-off area 0.39 nm$^2$/molecule. After that, the surface pressure rose smoothly up to 29.5 mN/m and get collapsed beyond this. The area per molecule corresponding to the collapse pressure was 0.21 nm$^2$. Considering molecule **2** as cuboid box with dimensions of three sides as 1.739 nm × 0.95 nm × 0.34 nm, three cross sectional area obtained as 1.65, 0.591 and 0.323 nm$^2$ (Supporting Information, *Figure S5*). These three areas correspond to three different possible orientation of molecule **2**. Limiting molecular area calculated from the isotherm curve is 0.313 nm$^2$, which is very close to 0.323 nm$^2$ i.e., cross sectional area of the molecules for vertical orientation. This suggests that the molecule **2** remained vertically orientated at air-water interface in Langmuir monolayer.

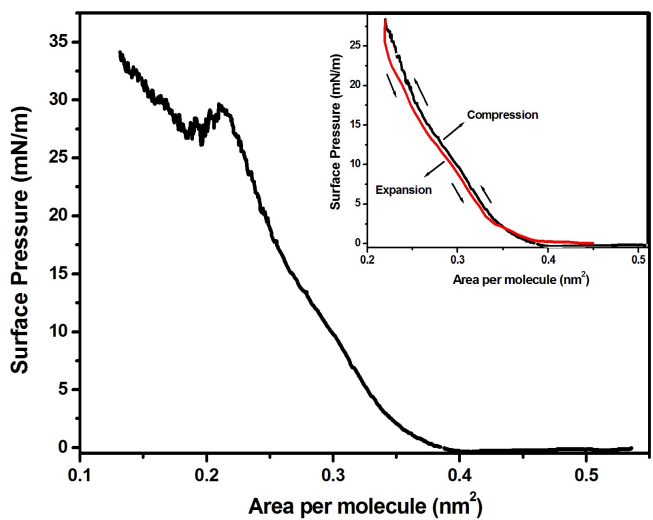

**Figure 2.** Surface pressure (π) Vs area per molecule (A) isotherm of **2**. Inset shows hysteresis of **2**.

During isotherm measurement successive compression-expansion cycles were also performed. The resulting isotherm of one successive compression and decompression experiments is shown in the inset of Figure 2. The result showed completely reversible behaviour with almost negligible shift of area per molecular during successive compression and expansion. It may be noted that for the molecules which do not form smooth stable film at air – water interface, such compression – expansion study shows clear hysteresis behaviour.[71,79] Also during isotherm measurement we have kept the barrier fixed at different surface pressures viz 15, 20, 25 mN/m for several hours. However, almost no change in surface pressure was observed even after 5 hours. This suggests molecule **2** form stable Langmuir film onto water surface. Therefore, observed characteristics from the π-A isotherm clearly supported the fact that molecule **2** formed almost ideal and stable monolayer at air-water interface.

## 2.2. Brewster angle microscopy (BAM) study

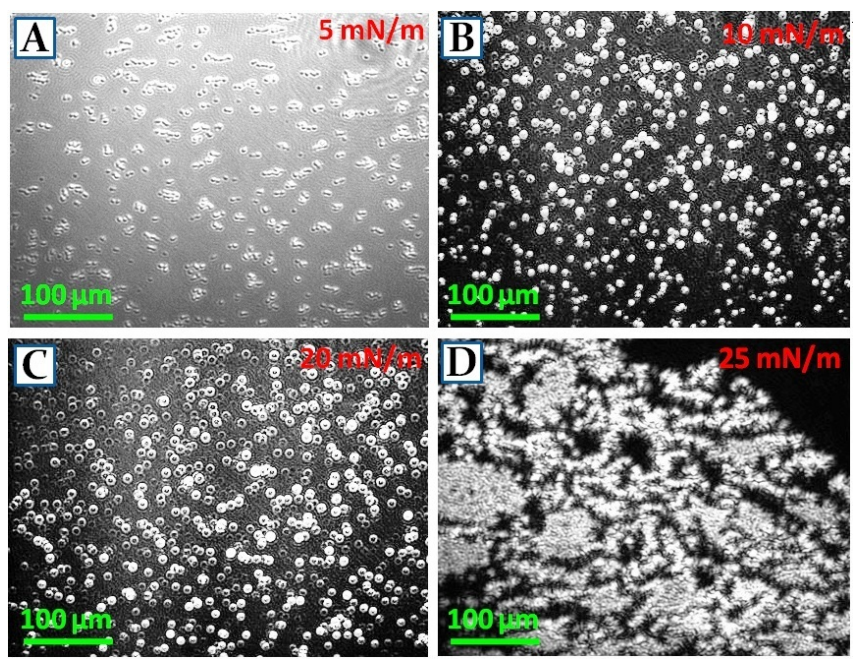

**Figure 3.** BAM images of **2** at surface pressure (A) 5 mN/m (B) 10 mN/m (C) 20 mN/m (D) 25 mN/m during π-A measurement at air-water interface.

Investigation of Langmuir film using BAM is one of the most suitable method to get visual information about the formation of domain structure of the floating film at water surface.[80] In the present case, BAM photographs for Langmuir film of **2** were taken at various surface pressures during π-A isotherm measurement. The corresponding BAM images are shown in Figure 3.

Here BAM photographs recorded at lower pressure (Figure 3A & 3B) revealed almost identical continuous small circular domains throughout the surface. The dimensions of the domains ranging from 5 μm–15 μm. However, at higher pressure (> 10 mN/m) marked changes (Figure 3C & 3D) in the BAM photographs were observed. Here, upon compression the number of domains and their structures in the BAM photographs changed gradually. At higher surface pressure the individual domains came close to each other and formed almost continuous and compact Langmuir film at water surface. As a whole BAM study gave compelling visual evidence about the formation of floating Langmuir film of **2** at water surface.

## 2.3. Atomic force microscopic (AFM) imaging of the transferred films

To have idea about the morphology and successful deposition of floating Langmuir film onto solid substrate, AFM studies have been performed. For the AFM study the floating Langmuir film of molecule **2** has been transferred onto smooth silicon wafer substrate at various surface pressures of 5, 10, 15, 20 mN/m. Representative AFM images (5 mN/m) with height profile are shown in Figure 4. In the later part of the manuscript the switching behaviour of compound **2** assembled onto LB film deposited at 15 mN/m surface pressure have been studied. Accordingly AFM images of the LB film deposited at 15 mN/m are also shown in *Figure S2* of Supporting Information.

AFM images revealed that the floating Langmuir film of molecule **2** have been successfully transferred onto solid substrate. All the AFM images irrespective of deposition pressure showed almost similar surface morphology. Surface coverage was more than 80%. Height profile analysis revealed that the height of the film lies within 4 – 8 nm range. On the other hand approximate length of the compound **2** is 3.7 nm (Supporting Information, *Figure S1*). Isotherm studies also suggested that the compound **2** remained vertically oriented with alkyl chain outward at air-water interface. If the compound **2** remained vertically oriented in solid substrate also, then it was expected that the maximum height should be 3.7 nm or less. Therefore, observed height profile clearly suggested that molecule **2** formed aggregated structures in certain places within the film when transferred onto solid substrate. Visual appearance of the AFM image also indicated the same.

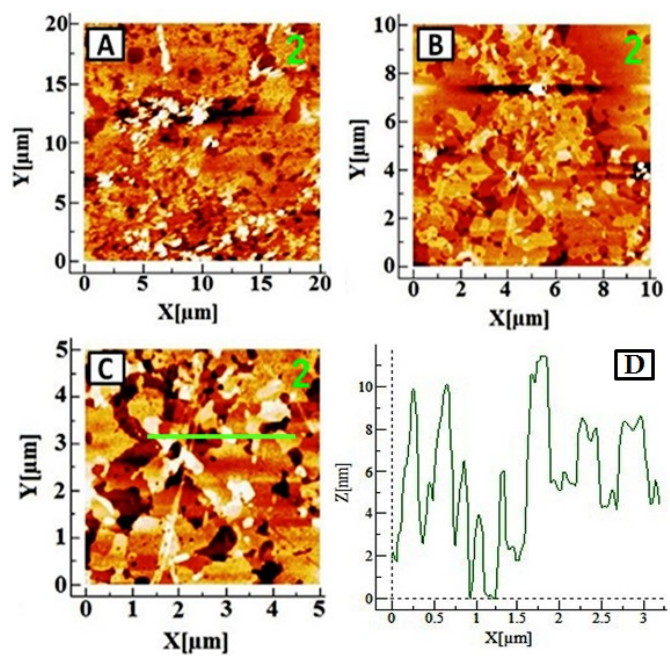

**Figure 4.** AFM image of **2** in the one-layer LB film deposited at 5 mN/m at different scan areas along with profile analysis spectra.

## 2.4. Resistive Switching Behaviour

Molecule **2** contains electron donor-acceptor groups and π-electron clouds (Figure 1). These are the indicative that molecule **2** might show electrical switching behaviours under suitable condition. Accordingly we have measured the I-V characteristics of molecule **2** deposited onto LB films. Here, we have used 60 layer LB films of molecule **2** to investigate the switching behaviour of molecule **2** having configuration Au/**2** LB film/ITO. Here **ITO** served as bottom electrode and **Au** acted as top electrode. The configurations of the device is shown in Supporting Information (*Figure S3*).

To study the switching behaviour of this device I–V characteristics was recorded by applying scanning voltage in two sweep directions which starts from $+V_{max}$ to $-V_{max}$ (forward scan) and $-V_{max}$ to $+V_{max}$ (reverse scan). The corresponding I-V characteristics of Au/**2** LB film/ITO device for scan range +2.5 V to -2.5 V is shown in Figure 5A.

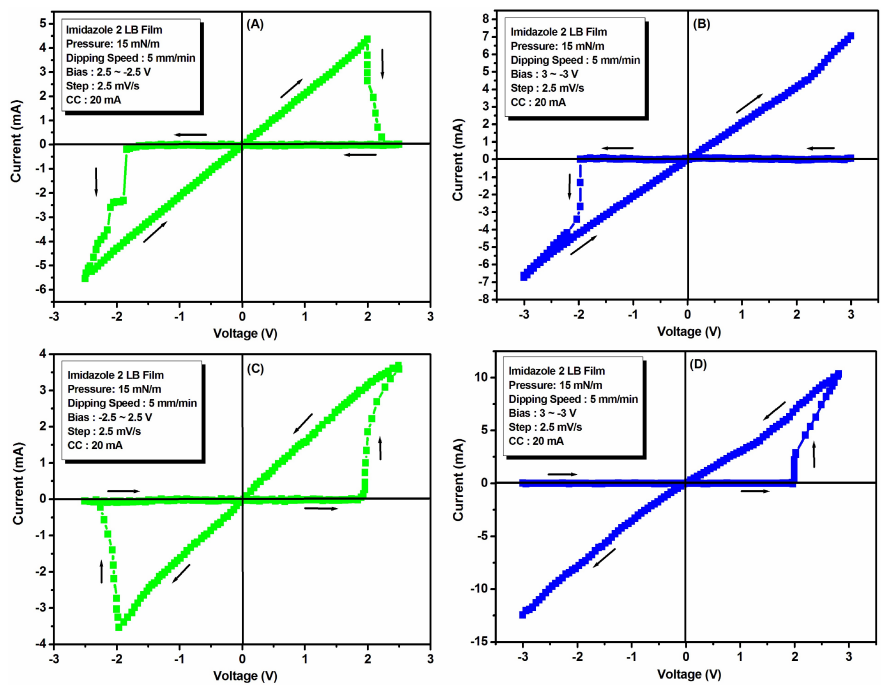

**Figure 5.** I-V characteristic for bipolar resistive switching of the Au/**2** LB Film/ITO device for two sweep direction. Initial sweep voltages were (A) ± 2.5 V, (B) ± 3 V. (C) ± 2.5 V and

(D) ± 3 V were applied in reverse sweep sequence. Arrows show the sweep direction of applied voltage.

Figure 5A revealed that during scaning from $+V_{max}$ to $-V_{max}$ at first the device exhibited its low conducting OFF state. The device switched from OFF state to ON state when the scanning voltage reached at -1.95 V (threshold voltage). When the scanning voltage is reversed ($-V_{max}$ to $+V_{max}$), the device again switched from its ON to OFF state at the threshold voltage +1.95 V. During measurement scaning voltage range has also been changed by increasing $V_{max}$ with an increment of 0.1 V in each time and almost similar bipolar switching behaviour was observed upto $V_{max}$ = 2.9 V. Although, when $V_{max}$ exceeded + 3 V once the device switcheed to its high conducting ON state (scaning from $+V_{max}$ to $-V_{max}$), it did not come back to its low conducting OFF state when scaning voltage is reversed (Figure 5B).

A reverse sweep sequence starting from $-V_{max}$ to $+V_{max}$ was also applied to Au/**2** LB film/ITO device to check the dependence of the switching behavior on the polarity of the first applied voltage sweep (Figure 5C). Results revealed, that here also the device switched from OFF state to ON state during the positive part of the voltage sweep (+ 1.92 V) and switched back from ON state to OFF at threshold voltage (-1.96 V) during the negative sweep. Like the previous one here also the device did not come back to its OFF state from ON state when $V_{max}$ exceeds 3 V (Figure 5D).

In order to check the scan rate dependence of the observed switching behaviors we have checked the switching by varying scanning speed. To realize this during I–V characteristic measurements, the sweep speed of the applied voltage was varied from 2.5 to 25 mV/s (Figure not shown). However, no significant change (threshold voltage, short circuit voltage etc.) were observed in the switching behaviour of compound **2** in LB films. This suggests that observed bipolar switching behaviour of compound **2** is scan rate independent.

I-V characteristic of molecule **2** in LB films with passage of time up to 10 days have also been studied. Almost identical bipolar switching behavior without any change in threshold voltage was observed even after 10 days. These results revealed that stability of the films with respect to switching behaviour was very good even after 10 days of film formation.

## 3. Discussion

Full understanding of the mechanism and underlying physics behind the switching phenomenon in the devices based on organic molecules are yet to be fully understandable. A substantial amount of research has been dedicated to the understanding of such switching phenomenon.[80–87] Based on theoretical simulation and experimental results several mechanism and strategies have been reported in order to explain such switching behaviour. These are reduction–oxidation process, conformational change, rotation of functional group, charge transfer, filamentary conduction, space charge and traps, ionic conduction, electron tunnelling and hopping etc.[81–88]

In general the electron acceptor group present in the organic molecule attracts the π electron clouds. This affects the conjugation within the molecule to a large extent resulting very low OFF state leakage current. Depending on the resulting built-in internal field due to difference in the work functions of the metal electrodes used as well as bias amplitude and bias direction, electron injection from electrode to the LUMO level of the organic layer (molecule) occurred. This results the device to switch into its high conducting ON state via electro-reduction. As a whole the high conducting ON state has been attained to be due to the modification of conjugation via electro-reduction of the molecules within the device. Therefore, based on the bias condition, conjugation within the molecular backbone gets extended and HUMO-LUMO gap reduced resulting the device in its high conducting ON state. Electro-reduction of the molecules in such devices may be due to, -

(i) electron injection overcoming the energy difference between the electrode work function and LUMO level.

(ii) hopping transport depending on applied field.

Again the device will return to its high resistance OFF state depending on the bias condition where the molecules get oxidized.

Reduction-oxidation process within the molecule may be responsible for the observed bipolar memory switching. C=N group (electron aceptor) present in the molecule **2** attracts the π-electron clouds towards them. This affects the conjugation in the molecule to a large extent –resulting very low OFF state leakage current. During scanning from $+V_{max}$ to $-V_{max}$ when the device is reverse biased to particular value (-1.95V) electron injection occurred from electrode to molecule. Here, low value of barrier height favours electron injection from ITO electrode to the LUMO level of molecule **2** (Figure 6a). At this bias voltage molecules near the ITO electrode are reduced. Accordingly electron density of the benzene ring rearranges and restore the electron conjugation in molecule **2**. Accordingly extension of the π electron clouds over the entire molecular surface occurs and facilitates a conduction pathway to switch the molecule in its high conducting ON state. Again when the voltage approaches to a suitable forward bias voltage during scanning from $-V_{max}$ to $+V_{max}$, the molecule get oxidized by withdrawing the previously added extra electron and the device comes back to its low conducting OFF state

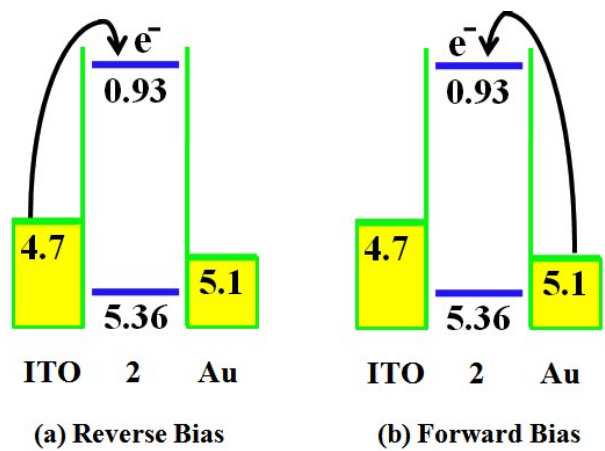

**Figure 6.** Energy level diagram of ITO/**2** LB Film/Au configuration under (a) reverse bias and (b) forward bias.

On the other hand when the device is scanned with a reverse sweep sequence starting from -$V_{max}$ to +$V_{max}$, the device also switches to its high conducting ON state at a particular value of forward voltage. Here, the Au electrode donates electron to the molecule **2**. Accordingly the molecules near the Au electrode get reduced. However, due to larger barrier height here electrons reaches the LUMO level of molecule **2** from Au electrode via hopping transport [42,48] (Figure 6b). This results the device in its ON state. Again the device come back to its low conducting OFF state at a particular reverse voltage while scanning from +$V_{max}$ to –$V_{max}$, where oxidation of the molecules occurred.

Different switching parameters as extracted from Figure 5A & 5C are listed in Table 1. A close look to the calculated values shows that the switching device shows higher current ON-OFF ratio as well as lower ON state resistance during scanning from + $V_{max}$ to – $V_{max}$. This is because at this sweep direction electron injection from ITO electrode to LUMO level occurred and the electron has to cross lower barrier height. On the other hand for other sweep direction electron reaches LUMO level from Au electrode via hopping transport and has to overcome larger barrier height. Accordingly current across the device is high during sweep direction from + $V_{max}$ to – $V_{max}$.

| Sweep Direction | OFF state Resistance (Ω) | On State Resistance (Ω) | Current ON-OFF Ratio |
|---|---|---|---|
| + V $_{Max}$ to – V $_{Min}$ (Fig.6A) | 34.5 x 10$^4$ | 46 x 10$^1$ | 794 |
| - V $_{Min}$ to + V $_{Max}$ (Fig.6C) | 33 x 10$^4$ | 55 x 10$^1$ | 591 |

**Table 1.** Different switching parameters for Au/2 LB Film/ITO device.

Also during scanning when V$_{max}$ exceeds 3 V, once the device switches to its high conducting ON state it does not return to its low conducting OFF state during scanning with reverse polarity (Figure 5B & 5D). This is true irrespective of the first applied bias voltage polarity. This observed unidirectional switching behaviour may be due to the short circuit of the device.[89] Here the organic layer gets heated due to passage of current through the device. When the current crosses certain limiting value due to increase in bias voltage, the organic layer within the device damaged and short circuit of the device occurred.

It is worthwhile to mention in this context that it has been observed that another imidazole derivative compound **1** showed threshold switching behaviour in LB films. Charge transfer between donor to acceptor was the key to such observed switching behaviour.[71] However, no bipolar switching was observed. Interestingly, chemical structure of molecule **2** used in the present study was almost identical to that of molecule **1**. The only difference is that in case of molecule **2** the – H group of NH of molecule **1** is replaced by a benzyl group. Here for molecule **2** bipolar switching is observed. This observed bipolar switching is due to reduction – oxidation process depending on the applied bias. Upon application of suitable bias electron are injected from electrode to the LUMO level of **2**. Due to the presence of strong reducible group (C = N) the molecules near the electrode get easily reduced. Accordingly conjugation in the molecule is restored and the device will switch ON to its high conducting state. Again during scanning towards opposite direction, at a specific bias voltage the molecules get oxidized leading to its low conducting OFF state. This situation has been shown in Figure 7. Whereas, presence of oxidizable group (N – H) in molecule **1** opposes the reduction process upon application of bias. Accordingly in case of molecule **1** no bipolar switching was observed. Therefore reduction – oxidation process in addition to intra-molecular charge transfer in case of molecule **2** play the key rule towards observed bipolar memory switching.

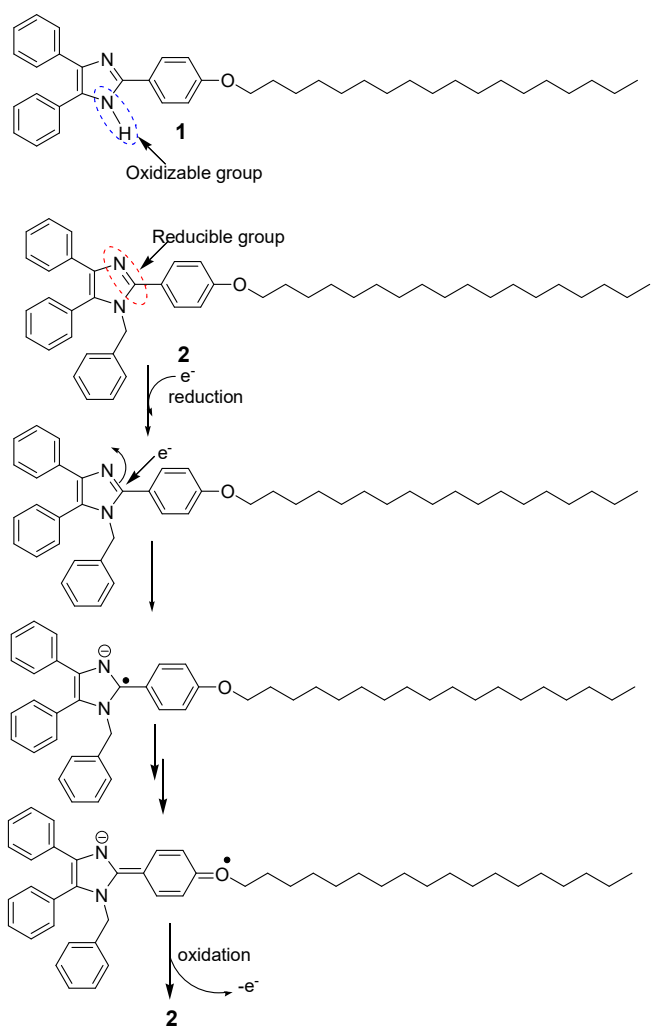

**Figure 7.** Reduction – oxidation steps involved in **2** during applied bias.

To confirm our assumption about reduction – oxidation process we have also performed cyclic voltammetry (CV) measurements for both the molecules **1** and **2** to study their oxidation and reduction behaviour (Supporting Information, *Figure S4*). Such measurements determine the potentials at which a single molecule is oxidized or reduced. Although, normally these values do not match with the critical voltages at which the organic material switches between the ON and OFF state.[83,90]

Our CV measurement clearly revealed that molecule **2** possessed prominent reduction potential peak at 0.17 V and oxidation potential peak at 0.3 V (Supporting Information, *Figure S4*). However, in case of molecule **1** no such sharp reduction – oxidation peaks was observed. As a whole presence of sharp reduction – oxidation peaks in the CV curve of **2** supported our assumptions of reduction – oxidation behaviour of molecule **2** during switching performance.

## 4. Conclusion

In conclusion, in the present study we have designed and synthesized an imidazole derivative compound **2** and studied its switching behaviour assembled onto LB films. Pressure – area ($\pi$ – A) isotherm, hysteresis analysis as well as BAM studies confirmed the formation of stable Langmuir film at water surface using molecule **2**. AFM studies gave visual evidence of the successful transformation of the floating film of molecule **2** onto solid substrate. I – V measurement of the device fabricated using 60 layer LB film of **2** showed bipolar switching behaviour at a threshold voltage of around 1.95 V. Reduction – oxidation process played crucial rule for such observed bipolar switching. Prominent reduction and oxidation peaks in the CV measurement also support this hypothesis. Presence of benzyl group in the imidazole core play vital rule in the reduction – oxidation process and hence to the observed bipolar memory switching behaviour of molecule **2**. When the applied bias voltage exceeded ±3 V the device got short circuited. Such switching behaviour of organic molecule is very important for realization of future organic electronics – specially for memory application and logic element in integrated circuit.

**Experimental Section**

**Materials:** 2-(4-O-octadecyloxyphenyl)-4,5-diphenyl imidazole (**1**) was synthesized according to the procedure described in our earlier report.[71] Dimethylformamide (DMF), anhydrous potassium carbonate and benzyl bromide were purchased from Fisher Scientific. Diethyl ether and anhydrous sodium sulphate were purchased E-merck India. Stock solutions were prepared by dissolving them in spectroscopic grade chloroform (SRL, India). DMF was dried and distilled prior to use. Chromatographic solvents were distilled before use. Other materials were used as received.

**Synthesis of 1-benzyl-2-(4-o-octadecyloxyphenyl)-4,5-diphenyl imidazole (2).**[71]

A solution of **1** [2-(4-*O*-octadecyloxyphenyl)-4,5-diphenyl imidazole, 0.75 mmol] in dry DMF (5 mL) was stirred under inert atmosphere with anhydrous $K_2CO_3$ (1 mmol) and benzyl bromide (0.90 mmol) at 60°C until the disappearance of the starting material (~ 4h). Solid precipitate formed was filtered and thoroughly washed with diethyl ether (25 mL). After extractive workup, the combined organic layer was dried over $Na_2SO_4$ and concentrated under reduced pressure. Purification of the crude product by column chromatography over silica-gel (60-120 mesh) using 2-5% ethyl acetate-hexane afforded title compound **2** as colourless solid. Yield 89%, m. p. 91°C; IR (KBr) $\nu_{max}$ 3020, 2941, 2856, 1597, 1467, 1369, 1270 cm$^{-1}$; $^1$H NMR (400 MHz, CDCl$_3$) δ 7.59 (m, 4H), 7.39-7.30 (m, 3H), 7.24-7.20 (m, 7H), 7.17-7.14 (m, 1H), 6.93 (d, *J* = 8.8 Hz, 2H), 6.85-6.83 (m, 2H), 5.11 (s, 2H), 3.98 (t, *J* = 6.4 Hz, 2H), 1.83-1.76 (m, 2H), 1.46-1.44 (m, 2H), 1.28 (s, 28H), 0.90 (t, *J* = 6.4 Hz, 3H); $^{13}$C NMR (100 MHz, CDCl$_3$) δ 159.8, 148.0, 137.5, 131.1, 130.5, 129.7, 128.8, 128.6, 128.1,

127.4, 126.9, 126.4, 126.0, 114.6, 68.1, 48.3, 31.9, 29.7, 29.6, 29.43, 29.39, 29.2, 26.0, 22.7, 14.2; HRMS requires for ($C_{46}H_{58}N_2O + H^+$) 655.4627, found 655.4622. $^1H$ and $^{13}C$ NMR spectrum of compound **2** has been given in Supporting Information (*Figure S5 & S6*).

**Isotherm measurement and film formation**

For surface pressure vs area per molecule (π-A) isotherms measurement and LB films preparation, a commercially available Langmuir-Blodgett (LB) film deposition instrument (Apex 2006C, Apex Instruments Co., India) was used. To measure the isotherm and film formation 120 μl of chloroform solution of **2** (concentration 0.5 mg/ml) was spread onto the subphase of pure Milli-Q water (18.2 MΩ–cm) at room temperature with the help of a micro syringe. After waiting 15 min to evaporate the solvent, the barrier was compressed at a speed of 5 mm/min to measure the pressure−area (π – A) isotherm. Wilhelmy plate arrangement was used to measure the surface pressure.[91] In order to avoid measurement error every isotherm was repeated a number of times. The isotherm presented here is an average of three measurements. For electrical characterisation floating film was deposited onto cleaned ITO coated glass substrate and for AFM measurement the same has been transferred onto silicon wafer substrate. Y-type deposition with deposition speed 5 mm/min was used to transfer the Langmuir film onto solid substrate with a transfer ratio 0.98±0.02.

**Brewster angle microscopy (BAM)**

Brewster Angle Microscopy (model- nanofilm_EP4-BAM, Accurion) was used to confirm the formation of floating Langmuir film of molecule **2** at water surface. Details of the BAM technique has been repeated elsewhere.[42]

**Atomic force microscopy (AFM)**

AFM technique has been used to have visual confirmation of successful transfer of the floating film onto solid substrate with an AFM instrument (Model – Innova, Bruker AXS Pte Ltd.) Details of the AFM technique has been reported elsewhere.[74]

**I-V characteristic measurement**

Source meter (model-2401) form Keithly was used for I-V measurement. 60 layer of molecule **2** has been deposited onto ITO coated glass substrate for this purpose. Details of I-V measurement technique has been given elsewhere.[42]

**Cyclic voltammetry (CV) measurement**

CV measurements was carried out using a software-controlled three-electrode electrochemical workstation (CHI 660E, CH Instruments). All the measurements were done at ambient condition. Details about CV measurement has been given elsewhere.[92]

**Density functional theory (DFT) study**

The HOMO and LUMO energy values for molecule **2** have been calculated through density functional theoretical calculations using Gaussian-09 suite of software package.[93] The geometry optimized structure and HUMO-LUMO surfaces of molecule **2** are given in Supporting Information (*Figure S7 & S8*).

# Acknowledgement

The authors acknowledges the financial support from to DST, Govt. of India, to carry out this research work through DST project Ref: EMR/2014/000234, Ref. No. SB/EMEQ-142/2014 and DST FIST project Ref: SR/FST/PSI-191/2014. The authors also acknowledges the financial support from UGC, Govt. of India, to carry out this research work through UGC –

SAP program Ref: No. F.530/23/DRS-I/2018 (SAP-I). Authors are very much thankful to Dr. G.G. Khan, Assistant Professor, Department of Material Science and Engineering, Tripura University for giving the opportunity to use the CV measurement facility in his laboratory. The authors also acknowledge experimental support offered by central instrumentation centre (CIC) Tripura University.

**Key words:** Organic electronics, Bipolar switching, Imidazole **2**, LB films, reduction – oxidation.

**For Table of Contents Use Only**

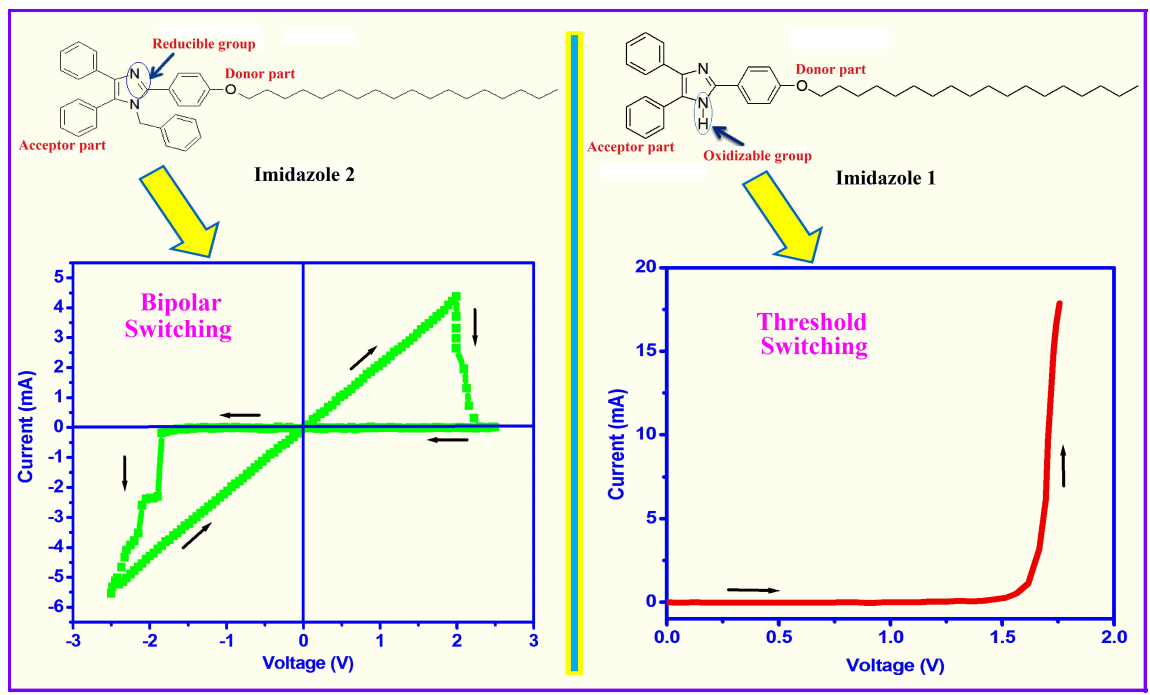

**Switching behaviour of imidazole derivative**

**Graphical abstract**